%% file: main.tex
\documentclass[11pt,times,mathptm,twoside,twocolumn,epsf]{article} 
\usepackage{times,fullpage}
\usepackage{graphicx,epsfig,url}
\usepackage[bf]{subfigure}
\usepackage{color}

\begin{document}
\newcommand{\rednote}[1]{\textcolor{red}{ \em #1 }}
\newcommand{\wendi}[1]{\footnote{{\bf Wendi: #1}}}

\input{header}

\input{abstract}

\input{intro}

\input{motivation}

\input{related-work}
\input{protocols}
\input{results}

\input{conclude}
\input{future-work}
\bibliography{sensor-nets,ad-hoc}
\bibliographystyle{ieee}
\end{document}

%% file: header.tex
\title{Collaborative Storage Management In Sensor Networks}
\author{Sameer Tilak, Nael B. Abu-Ghazaleh and Wendi Heinzelman}
\author{Sameer Tilak$^\dag$, Nael B. Abu-Ghazaleh$^\dag$ and Wendi
Heinzelman$^\ddag$\\
         \\
\begin{minipage}{80mm}
\begin{center}
         $^\dag$Computer System Research Laboratory \\
         Dept. of CS, Binghamton University \\
         Binghamton, NY~~13902--6000 \\
         \url{{sameer,nael}@cs.binghamton.edu} 
  \end{center}
 \end{minipage}\begin{minipage}{80mm}
 \begin{center}
         $^\ddag$Dept. of Electrical and Computer Engr.\\
         University of Rochester  \\
         Rochester, NY~~14627--0126 \\
         \url{wheinzel@ece.rochester.edu} 
 \end{center}
\end{minipage}
}
\maketitle

%% file: abstract.tex
\begin{abstract}

 In this paper, we consider a class of sensor networks where the data
 is not required in real-time by an observer; for example, a sensor
 network monitoring a scientific phenomenon for later play back and
 analysis. In such networks, the data must be stored in the network.
 Thus, in addition to battery power, storage is a primary resource:
 the useful lifetime of the network is constrained by its ability to
 store the generated data samples. We explore the use of collaborative
 storage technique to efficiently manage data in storage constrained
 sensor networks.  The proposed collaborative storage technique takes
 advantage of spatial correlation among the data collected by nearby
 sensors to significantly reduce the size of the data near the data
 sources.  We show that the proposed approach provides significant
 savings in the size of the stored data vs. local buffering, allowing
 the network to run for a longer time without running out of storage
 space and reducing the amount of data that will eventually be relayed
 to the observer.  In addition, collaborative storage performs load
 balancing of the available storage space if data generation rates are
 not uniform across sensors (as would be the case in an event driven
 sensor network), or if the available storage varies across the
 network. 
\end{abstract}

%% file: intro.tex
\section{Introduction}
 Wireless Sensor Networks (WSNs) hold the promise of revolutionizing 
 sensing across a range of civil, scientific, military and industrial 
 applications.  
 However, many battery-operated sensors have constraints such as
 limited energy, computational ability, and storage capacity, and thus
 protocols must be designed to deal efficiently with these limited
 resources.

In this paper, we consider a class of sensor networks where the
information collected by the sensors is not collected in real-time.
In such applications, the data must be stored, at least temporarily,
within the network until it is later collected by an observer (or
until it ceases to be useful).  Such applications include scientific
monitoring: the sensors are deployed to collect detailed information
about a phenomenon for later playback and analysis.  In addition, some
applications have sensors which collect data that may be needed by
users of the networks that generate queries dynamically.  In such
applications, the data must be stored in the network; storage becomes
a primary resource which, in addition to energy, determines the useful
lifetime of the network.  This paper considers the problem of storage
management in such networks: how to use limited persistent storage of
a sensor to store sampled data effectively.
In addition to the applications above, storage can be used to tolerate
temporary network partitioning, where the observer is not reachable
from the partitioned sensors, without losing potentially valuable
data.

One basic storage management approach is to buffer the data locally at
the sensors that collect them.  However, such an approach does not
capitalize on the spatial correlation of data among neighboring
sensors to reduce the overall size of the stored data (the property
that makes data aggregation possible~\cite{intanagonwiwat-01}).
Collaborative storage management can provide the following advantages
over a simple buffering technique: (1) More efficient storage allows
the network to continue storing data for a longer time without
exhausting storage space; (2) Load balancing is possible: if the rate
of data generation is not uniform at the sensors (e.g., in the case
where a localized event causes neighboring sensors to collect data
more aggressively), some sensors may run out of storage space while
space remains available at others. In such a case, it is important for
the sensors to collaborate to achieve load balancing for storage to
avoid or delay data loss due to insufficient local storage; and
(3) Dynamic, localized reconfiguration of the network (such as
adjusting sampling frequencies of sensors based on estimated data
redundancy and current resources).

We describe a cluster-based collaborative storage approach and compare
it through simulations to a local buffering technique. Our experiments 
show that collaborative storage makes more efficient use of sensor storage 
and provides load balancing, especially if a high level of spatial
correlation among the data of neighboring sensors is present.  The
trade-off is that using collaborative storage, data need to be
communicated among neighboring nodes, and thus collaborative storage
expends more energy than local buffering. However, since data is
aggregated using collaborative storage, a smaller amount of data is
stored and a smaller amount of data is eventually relayed to the
observer, thereby reducing energy dissipation in this phase of
operation. 
We then explore the use of coordination for redundancy control.  More
 specifically, the cluster head can evaluate the amount of redundancy
 present among neighboring sensors, and use feedback this information
 back to the sensors to adjust their sampling rate.
We exploit coordination in conjunction with local storage as well as
collaborative storage and show that it provides desirable properties
in both cases.

The remainder of this paper is organized as follows.
Section~\ref{application} overviews the partitioned sensor network
problem and motivates collaborative storage in more detail in the
context of this problem.  Section~\ref{related} provides an overview
of related work in this area.  
Section~\ref{protocols} presents the
proposed storage management protocols and discusses the important
design tradeoffs.  In section~\ref{experimental} we evaluate the
storage alternatives under different scenarios.  Finally
Section~\ref{conclude} presents conclusion and our future
research.

%% file: motivation.tex
\section{Motivation}~\label{application}

In this section,
we describe two different applications which require in-network
storage. ZebraNet~\cite{juang-02}, is a sensor network application for
wild-life tracking whose goal is to provide more insight into complex
issues such as migration patterns, social structures and mobility
models of various animal species. In this application, sensors are
attached to animals.  Scientists (aka observers) collect the data by
driving around the monitored habitat receiving information from Zebras
as they come in range with them.  Data collection is not preplanned:
it might be unpredictable and infrequent.  The sensors do not have an
estimate regarding the observer's schedule. 
The observer would like the network to maintain all the new data
samples available since the last time the data was collected.
Further, we would like the collection time to be small since the
observer may not be in range with the zebra for a long time.

The second example application is a Remote Ecological Micro-Sensor
Network~\cite{remote-eco} aimed at remote visual surveillance of
federally listed rare and endangered plants.  This project aims to
provide near-real time monitoring of important events such as
visitation by pollinators and consumption by herbivores 
along with monitoring a number of weather conditions and events.
Sensors are placed in different habitats, ranging from scattered low
shrubs to dense tropical forests.  Environmental conditions can be
severe; e.g., some locations frequently freeze.
In this application, network partitioning (relay nodes becoming
unavailable) may occur due to the extreme physical conditions (e.g.,
deep freeze).
Important events that occur during disconnection periods should be
recorded and reported once the connection is reestablished. Effective
storage management is needed to maximize the partitioning time that
can be tolerated.

%% file: related-work.tex
\section{Related Work}\label{related}

Because of the wireless nature of sensors, the primary resource
constraint is the limited battery energy available.  Energy-awareness 
permeates all aspects of sensor design and operation, from the
physical design of the sensor~\cite{asada-98,chandrakasan-99} to the
design of its operating system~\cite{hill-00}, communication protocols
and applications~\cite{yao-02}.

Ratnasamy et al.~propose using Data Centric Storage (DCS) to store
data by name within a sensor network such that all related data is
stored at the same (or nearby) sensor nodes using geographic
hashing~\cite{ratnasamy-02}.  Thus, queries for data of a certain type
are likely to be satisfied by a small number of nodes, significantly
improving the performance of queries.  However, this enhanced query
performance requires moving related data from its point of generation
to its appropriate keeper as determined by geographic hashing.  We
view this work as a higher level management of data focusing on
optimizing queries rather than storage: our approach could compliment
DCS by providing more effective storage of the data as it is
collected.

Concurrently with us~\cite{tilak-icnp-03}, Ganesan et al have explored
protocols for storage constrained sensor networks~\cite{Ganesan03a}.
The work by Ganesan et al considers the same problem and explores some
of the solution space we are considering.  Our work differs in the
following ways: (1) We explore additional approaches to storage
management, including those using coordination; (2) we explore issues
that arise due to uneven data generation (e.g., due to event driven,
or adaptive sampling applications), and non-uniform storage
distribution (e.g., due to non-uniform deployment of the sensors).  In
such applications, effective load balancing is required; and (3) we
study some additional characteristics of the storage protocols
including coverage and collection time and energy.

%% file: protocols.tex
\section{Storage Management Protocols}~\label{protocols}
A primary objective of storage management protocols is to efficiently
utilize the available storage space to continue collecting data for
the longest possible time without losing samples in an energy efficient 
way.
Storage management approaches can be classified as:
\begin{enumerate}
\item \emph{Local storage}: This is the simplest solution where every
sensor stores its data locally. This protocol is energy efficient
during the storage phase since it requires no data communication.
Even though the storage energy is high (due to all the data being
stored), the current state of technology is such that storage costs
less than communication.  However, this protocol is storage
inefficient since the data is not aggregated and redundant data is
stored among neighboring nodes. Local storage is unable to
load balance if data generation or the available storage varies
across sensors.

\item \emph{Collaborative storage}: Collaborative storage refers to
  any approach where nodes collaborate.  Collaboration leads to two
  benefits: (1) Less data is stored: measurements obtained from nearby
  sensors are typically correlated.  This allows data samples from
  neighboring sensors to be aggregated; and (2) Load balancing:
  collaboration among sensors allows them to load balance storage.
\end{enumerate}

\noindent
It is important to consider the energy implications of collaborative
storage relative to local storage.  Collaborative storage requires
sensors to exchange data, causing them to expend energy during the
storage phase.  However, because they are able to aggregate data, the
energy expended in storing this data to a storage device is reduced.
In addition, once connectivity with the observer is established, less
energy is needed during the collection stage to relay the stored data
to the observer.  
We note that this holds true even if in-network aggregation is carried
out for locally buffered data during the reach-back stage due to the
following two reasons: (1) Initial communication (first hop) of the
locally buffered data will not be aggregated; and (2) Less efficient
aggregation: a smaller amount of time and resources is available when
near real-time data aggregation is applied during reach-back as
compared to aggregation during the storage phase.  Aggregating data
during reach-back is limited because all the data collected during the
storage phase is compressed in a short time.

\subsection{Collaborative Storage Protocols}

Within the space of collaborative storage, a number of protocols are
possible.  The primary protocol we study is Cluster Based
Collaborative Storage (CBCS). CBCS uses collaboration among nearby
sensors only: these have the highest likelihood of correlated data and
require the least amount of energy for collaboration.  We did not
consider wider collaboration because the collaboration cost may become
prohibitive; the cost of communication is significantly higher than
the cost of storage under current technologies.  The remainder of this
section describes CBCS operation.

In CBCS, clusters are formed in a distributed connectivity-based or
geographically-based fashion -- almost any one-hop clustering
algorithm would suffice.  Each sensor sends its observations to the
elected Cluster Head (CH) periodically. The CH then aggregates the
observations and stores the aggregated data. Only the CH needs to
store aggregated data, thereby resulting in low storage.  The clusters
are rotated periodically to balance the storage load and energy usage.
Note that only the CH needs to keep its radio on during its tenure,
while a cluster member can turn off its radio except when it has data
to send. This results in high energy efficiency: idle power consumes
significant energy in the long run if radios are kept on.  The
reception of unneeded packets while the radio is on also consumes
energy.

Operation during CBCS can be viewed as a continuous sequence of rounds
until an observer/base station is present and the reach-back stage can
begin. Each round consists of two phases: (1) CH Election phase: 
In this phase, each sensor advertises its resources to its one hop 
neighbors. Based on this resource information a cluster head (CH) is selected
The remaining nodes then attach themselves to that CH during the data
transfer phase; and (2) Data exchange phase: If a node is connected to
a CH, it sends its observations to the CH; otherwise, it stores its
observations locally.

\noindent
The CH election approach used in CBCS is based on the characteristics
of the sensor nodes such as available storage, available energy or
proximity to the ``expected'' observer location. The criteria for CH
selection can be arbitrarily complex; in our experiments we used
available storage as the criteria.

There has been considerable research in cluster formation algorithms
for MANETs that considered both static and dynamic cluster head
election.  Our requirements for the clustering algorithm are that it
be light-weight and localized -- only one-hop clusters.  Moreover, we
require cluster head rotation for load balancing of energy and
storage.  This is an idea borrowed from the LEACH
protocol~\cite{heinzelman-phd}.  The approach we use is a
representative one and there is room for future improvements in this
aspect of the protocol.

CH rotation is done by repeating the cluster election phases with
every round. The frequency of cluster rotation influences the
performance of the protocol.  Depending on the cluster formation
criteria, there is an overhead for cluster formation due to the
exchange of messages.

The cluster election approach above may result in a situation where a
node A, selects a neighbor B to be its CH when B itself selects C
(which is out of range with A) to be its own CH.  This may result in
chains of cluster heads leading to ineffective/multi-hop clustering.
To eliminate the above problem and restrict clusters to one hop,
geographical zoning is used: an idea that is similar to the approach
of constructing virtual grids~\cite{xu-01}.  More specifically, the
sensor field is divided into zones such that all nodes within a zone
are in range with each other.  Cluster selection is then localized to
a zone such that a node only considers cluster advertisements
occurring in its zone.  Only one CH is selected per zone, eliminating
CH chaining as discussed above.  We note that this approach requires
either pre-configuration of the sensors or the presence of a location
discovery mechanism (GPS cards or a distributed localization
algorithm~\cite{bulusu-00}). In sensor networks, localization is of
fundamental importance as the physical context of the reporting
sensors must be known in order to interpret the data. We therefore
argue that our assumption that sensors know their physical
co-ordinates is realistic.

\subsection{The Role of Coordination}

One idea we explore is coordination among the sensors.  Specifically,
each sensor has a local view of the phenomenon, but cannot assess the
importance of its information given that other sensors may report
correlated information.  For example, in an application where 3
sensors are sufficient to triangulate a phenomenon, 10
sensors may be in a position to do so and be storing this information
locally or sending it to the cluster head for collaborative storage.
Through coordination, the cluster head can inform the nodes of the
degree of the redundancy allowing the sensors to alternate
triangulating the phenomenon.  Coordination can be carried out
periodically at low frequency, with a small overhead (e.g., with CH
election).  Similar to CH election, the nodes exchange meta data
describing their reporting behavior and we assume that some
application specific estimate of redundancy is performed to adjust the
sampling rate.

As a result of coordination, it is possible that a significant
reduction in the data samples produced by each sensor is achieved.  We
note that this reduction represents a portion of the reduction that is
achieved from aggregation.  For example, in a localization
application, with 10 nodes in position to detect an intruder, only 3
nodes are needed.  Coordination allows the nodes to realize this and
adjust their reporting so that only 3 sensors produce data in every
period.  However, the three samples can still be aggregated into the
estimated location of the intruder once the values are combined at the
cluster head.

Coordination can be used in conjunction with local storage or
collaborative storage.  In Coordinated Local Storage (CLS), the
sensors coordinate periodically and adjust their sampling schedules to
reduce the overall redundancy, thus reducing the amount of data that
will be stored.  Note that the sensors continue to store their
readings locally. Relative to Local Storage (LS), CLS results in a
smaller overall storage requirements and savings in energy in storing
the data.  This also results in a smaller and more energy efficient
data collection phase.  Similarly, Coordinated Collaborative Storage
(CCS) uses coordination to adjust the sampling rate locally.  Similar
to CBCS, the data is still sent to the cluster head where aggregation
is applied.  However, as a result of coordination, a sensor can adapt
its sampling frequency/ data resolution to match the application
requirements.
In this case, the energy in sending the data to the cluster head is
reduced because of the smaller size of the generated data, but the
overall size of the data is not reduced.  We evaluate CLS and CCS
compared to the non-coordinated counterparts, LS and CBCS.

%% file: results.tex
\section{Experimental Evaluation}~\label{experimental}

We simulated the proposed storage management protocols using the NS-2
simulator~\cite{ns-2}.  We use a CSMA based MAC layer protocol.  A
sensor field of $350 \times 350$ meters$^{2}$ is used with each sensor
having a transmission range of 100 meters.  We considered three levels
of sensor density: 50 sensors, 100 sensors and 150 sensors deployed
randomly.
We divide the field into 25 zones (each zone is $70 \times 70$
meters$^{2}$ to ensure that any sensor in the zone is in range with
any other sensor).
The simulation time for each scenario was set to 500 seconds and each
point represents an average over five different topologies.  Cluster
rotation and coordination are performed every 100 seconds in the
appropriate protocols.  

We assume sensors have a constant sampling rate (set to one sample per
second). Unless otherwise indicated, we set the aggregation ratio to a
constant value of 0.5.  For the coordination protocols, we used a
scenario where the available redundancy was on average 30\% of the
data size -- this is the percentage of the data that can be eliminated
using coordination.  We note that this reduction in the data size
represents a portion of the reduction possible using aggregation.
With aggregation the full data is available at the cluster head and
can be compressed at a higher efficiency.
Several sensor nodes that are appearing on the market, including
Berkeley MICA nodes~\cite{mica} have Flash memories.
Flash memories have excellent power dissipation properties
and small form factor.
As a representative we consider a SimpleTech flash memory USB
cards~\cite{flash} with as Transfer Energy/Mbyte $0.055 J$. 
In current wireless communication technologies (Radio Frequency
based), the cost of communication is high compared to the cost of
storage.  For example, representative radios following the Zigbee IEEE
802.15.4 standard consume energy at roughly 40 times the cost of the
SimpleTech USB card above per unit data.  Our energy models in the
simulation are based on these two devices.  Further, we adjust the
radio properties to match those of a Zigbee device.

Note that the possible data aggregation/compression as well as the
reduction due to coordination are application as well as topology
dependent. Consider a temperature sensing application.  For this
application a given sensor can collect data from all its neighbors and
then simply take the average and store a single value (or maybe
minimum, mean and maximum values) as representative.  However, if the
sensors are sending video data, then such high spatial compression
might not be possible.  In this paper, instead of considering a
specific application, we assume a data aggregation model where the
cluster head is able to compress the size of the data by an
aggregation ratio $\alpha$.  By controlling $\alpha$ we can consider
different applications with different levels of available spatial
correlation.  In this model, the size of the aggregated data grows
linearly with the number of available sensors.  We consider the
implications of this model on collaborative storage and explore other
possible models later in this section.  We would like to emphasize
that we have selected these numbers as just representatives to
illustrate the the various tradeoffs. Due to space constraints we can
not include the results comparing all these protocols with various
values of aggregation ratio.

\subsection{Storage and Energy Tradeoffs}
\begin{figure}[t]
\epsfig{file=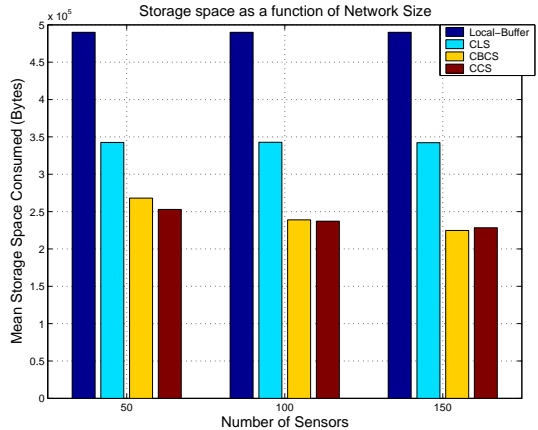,scale=0.4,silent=}
\caption{Storage space vs. Network Density}~\label{fig:all-storage}
\end{figure}

Figure~\ref{fig:all-storage} shows the average storage used per sensor
as a function of the number of sensors (50, 100 and 150 sensors) for
the four storage management techniques : (1) local storage (LS); (2)
Cluster-Based Collaborative Storage (CBCS); (3) Coordinated Local
Storage (CLS); and (4) Coordinated Collaborative Storage (CCS). 
In the case of CBCS aggregation ratio was set to 0.5.
The storage space consumption is independent of the density for LS and
is greater than storage space consumption than CBCS and CCS (roughly
in proportion to the aggregation ratio).  CLS storage requirement is in
between the two approaches because it is able to reduce the storage
requirement using coordination (we assumed that coordination yields
improvement uniformly distributed between 20\% and 40\%).  Note that
after data exchange, the storage requirement for CBCS and CCS are
roughly the same since aggregation at the cluster head can reduce the
data to a minimum size, regardless of whether coordination took place
or not.

Surprisingly, in the case of collaborative storage, the storage space
consumption decreases slightly as the density increases. While this is
counter-intuitive, it is due to higher packet loss observed during the
exchange phase as the density increases; as density increases, the
probability of collisions increases.  These losses are due to the use
of a contention based unreliable MAC layer protocol: when a node wants
to transmit its data to the CH. The negligible difference in the
storage space consumption between CBCS and CCS is also an artifact
slight difference in the number of collisions observed in the two
protocols.
The use of a reliable protocol such as that in IEEE 802.11 or a
reservation based protocol such as the TDMA based protocol employed by
LEACH~\cite{heinzelman-phd} can be used to reduce or eliminate losses
due to collisions (at an increased communication cost).  We leave the
exploration of these tradeoffs to future work.  Packet loss ranged
from around 1\% for the 50 sensor case to around 10\% for the 150
sensor scenarios.  Regardless of the effect of collisions, one can
clearly see that the collaborative storage achieves significant
savings in storage space compared to local storage protocols (in
proportion to the aggregation ratio).

\begin{figure}[t]
\epsfig{file=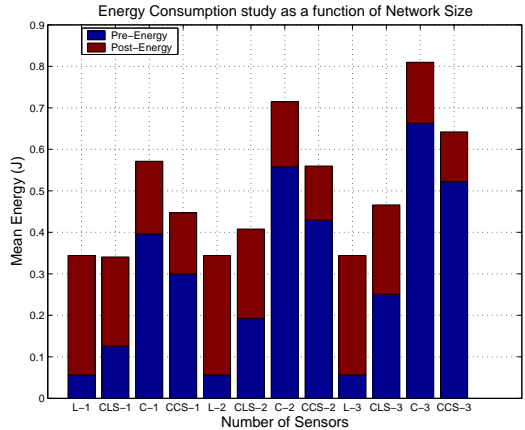,scale=0.4,silent=}
\caption{Energy consumption vs. Density}~\label{fig:all-energy}
\end{figure}

Figure~\ref{fig:all-energy} shows the consumed energy for the
protocols in Joules as a function of network density. 
The X-axis represents protocols for different network densities: L and
C stand for local buffering and CBCS respectively. L-1,L-2,and L-3
represents the results with local buffering technique for network size
50,100 and 150 respectively. The energy bars are broken into two
parts: pre-energy, which is the energy consumed during the storage
phase, and post-energy, which is the energy consumed during data
collection (the relaying of the data to the observer).  The energy
consumed during storage phase is higher for collaborative storage
because of the data communication among neighboring nodes (not present
in local storage) and due to the overhead for cluster rotation.
 CCS spends less energy than CBCS due to reduction in data size that
 results from coordination.  However, CLS has higher expenditure than
 LS since it requires costly communication for coordination.  This
 cost grows with the density of the network because our coordination
 implementation has each node broadcasting its update and receiving
 updates from all other nodes.  

For the storage and communication technologies used, the cost of
 communication dominates that of storage.  As a result, the cost of
 the additional communication during collaborative storage might not
 be recovered by the reduced energy needed for storage except at very
 high compression ratios.  This tradeoff is a function of the ratio of
 communication cost to storage cost; if this ratio goes down in the
 future (for example, due to the use of infra-red communication or
 ultra-low power RF radios), collaborative storage becomes more energy
 efficient compared to local storage.  Conversely, if the ratio goes
 up, collaborative storage becomes less efficient.

 The data collection model depends on the application and network
organization; several models are in use for deployed sensor networks.
We use a simple collection model where we only account for the cost of
transferring the data one hop.  This model is representative of an
observer that moves around and gather data from the sensors. Also, in
cases where the local buffering approach carries out aggregation at
the first hop towards the observer, the size of the data becomes
similar in the two approaches and the remainder of the collection cost
is the same.  However, this is slightly optimistic in favor of local
storage because near real-time data aggregation will not in general be
able to achieve the same aggregation level during collection as is
achieved during collaborative storage.  This is due to the fact that
collaborative storage can afford to wait for samples and compress them
efficiently.  Moreover, in collaborative storage, the aggregation is
done incrementally over time, requiring fewer resources than
aggregation during collection where large amounts of data are
processed during a short time period.  The collaborative storage
approaches outperform the local storage ones according to this metric
due to their smaller storage size.  CLS outperforms LS for the same
reason.

\begin{figure}
\epsfig{file=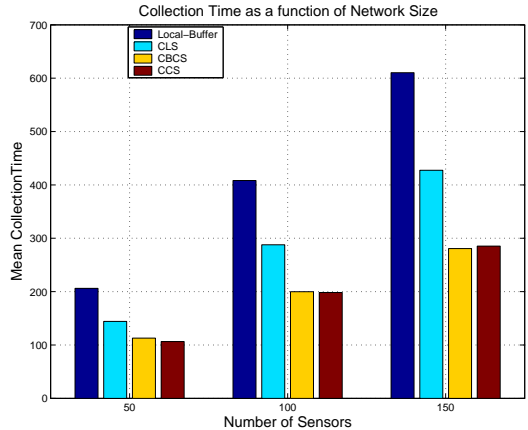,scale=0.4,silent=}
\caption{Mean collection time vs. Density}~\label{fig:ctim-nwsize}
\end{figure}

Figure~\ref{fig:ctim-nwsize} shows that with collaborative storage,
the collection time is considerably lower than that of local
buffering.  In addition, CLS outperforms LS.  Low collection time and
energy are important parameters from a practical standpoint.  After
exploring the effect of coordination, the remainder of the paper
presents results only with the two uncoordinated protocols (LS and
CBCS).

\subsection{Storage Balancing Effect}

In this study, we explore the load-balancing effect of collaborative
storage.  More specifically, the sensors are started with a limited
storage space and the time until this space is exhausted is
tracked.  We consider an application where a subset of the sensors
generates data at twice the rate of the others, for example, in
response to higher observed activity close to some of the sensors. To
model the data correlation, we assume that sensors within a zone have
correlated data. Therefore all the sensors within a zone will report
their readings with the same frequency. We randomly select zones with
high activity.; sensors within those zones will report twice as often
as those sensors within low activity zone.

\begin{figure}
\centerline{\epsfig{file=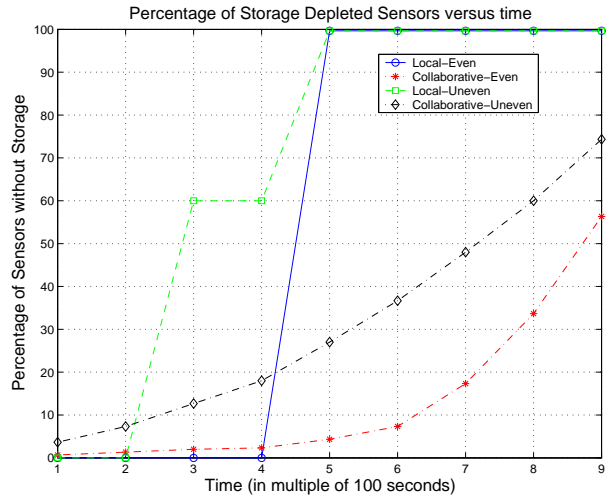,width=1.0\linewidth,silent=}}
\caption{Percentage of storage depleted sensors as a function of time.}~\label{fig:strless}
\end{figure}

In Figure~\ref{fig:strless}, the X-axis denotes time (in multiples of
100 seconds), whereas the Y-axis denotes the percentage of sensors
that have no storage space left.  Using LS, in the even data
generation case, all sensors run out of storage space at the same time
and all data collected after that is lost.  In comparison, CBCS
provides longer time without running out of storage because of its
more efficient storage.

The uneven data generation case highlights the load-balancing
capability of CBCS.  Using LS, the sensors that generate data at a
high rate exhaust their storage quickly; we observe two subsets of
sensors getting their storage exhausted at two different times.  In
comparison, CBCS has much longer mean sensor storage depletion time
due to its load balancing properties, with sensors exhausting their
resources gradually, extending the network lifetime much longer than
LS.

\subsection{Coverage Analysis}

  Physically co-located sensors have redundant data. For simplicity,
  we assume that all sensors within a zone have correlated data. In
  this work we consider two types of coverage, namely, binary coverage
  and manifold coverage, defined as follows: (1) Binary Coverage: A
  given zone $Z_{i}$ is said to be covered at time $t$ if any one of
  the sensors $S_{1} \ldots S_{k}$ in $Z_{i}$ is reporting and storing
  the reading. Binary coverage can be visualized as a step function;
  and (2) Manifold Coverage: A given zone $Z_{i}$ is said to be
  covered at time $t$ proportional to the number of sensors $j$ ($j <
  k$) out of its given set of sensors $S_{1} \ldots S_{k}$ that are
  reporting and storing the reading.  This coverage function can be
  visualized as a monotonically increasing function (which might have
  diminishing returns after some point).  This means that the higher
  the number of reporting sensors the better the coverage.

\begin{figure}
\centerline{\epsfig{file=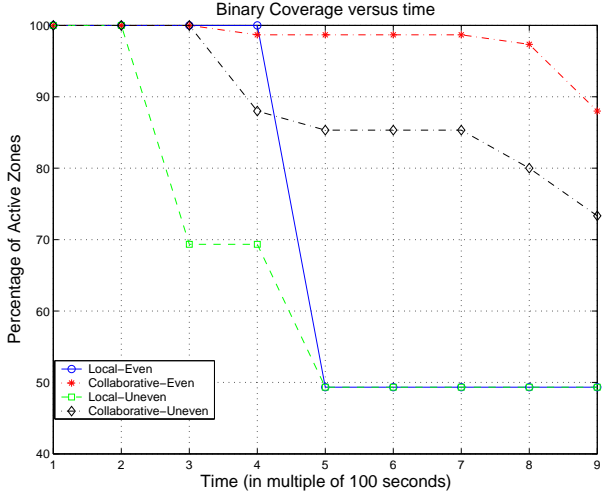,width=1.0\linewidth,silent=}}
\caption{Binary Coverage}~\label{fig:binarycoverage}
\end{figure}

Figure~\ref{fig:binarycoverage} shows the Binary Coverage as a
function of time.  One can see that CBCS has a higher percentage of
active zones compared to LS for both data generation models.

\begin{figure*}[t]
\begin{center}
\mbox{ \subfigure[LS Manifold Coverage (Even Data
Generation)\label{localmanieven}]{\epsfig{file=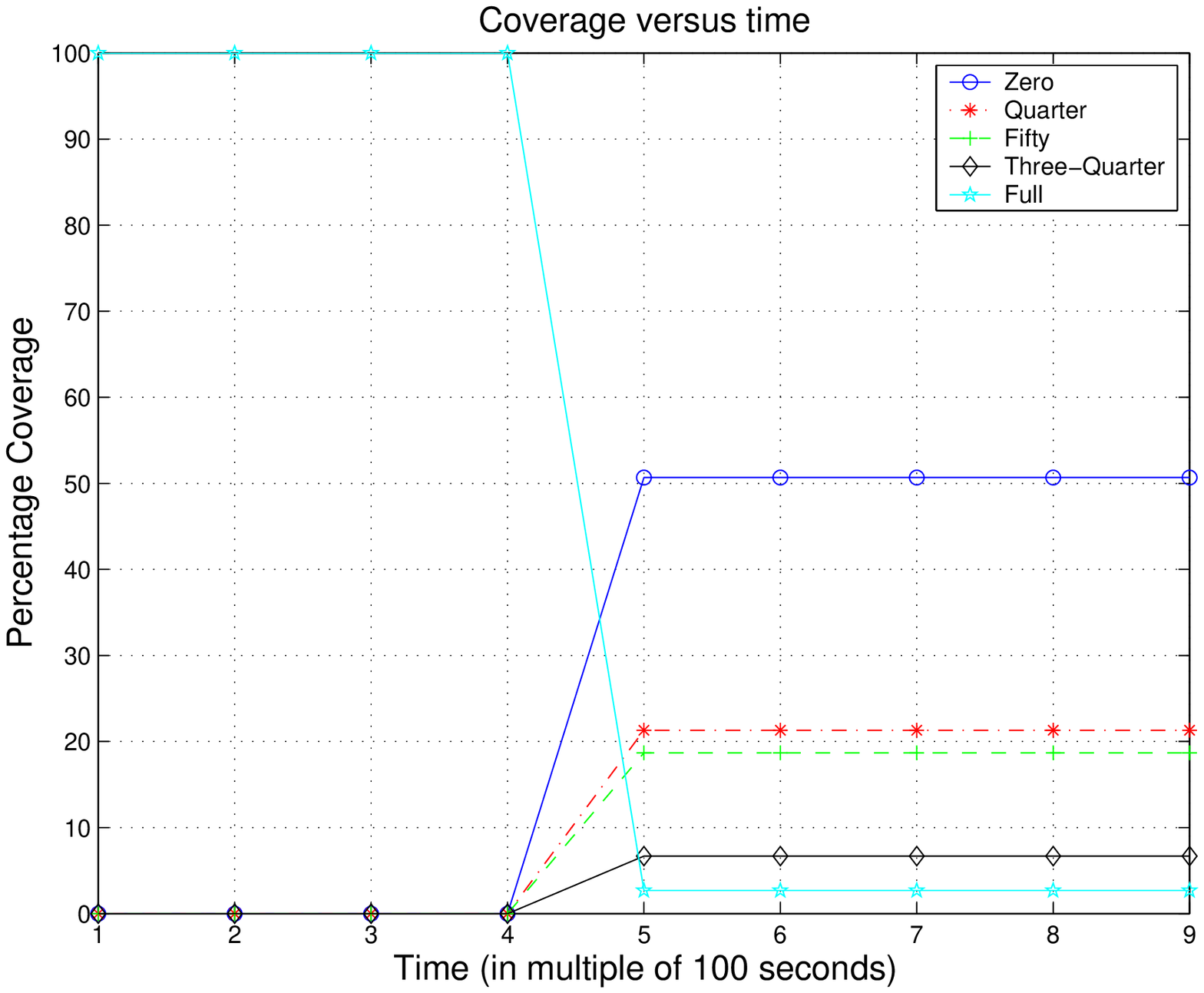,scale=0.4}}
\quad \subfigure[CBCS Manifold Coverage (Even Data
Generation)\label{clustmanieven}]{\epsfig{file=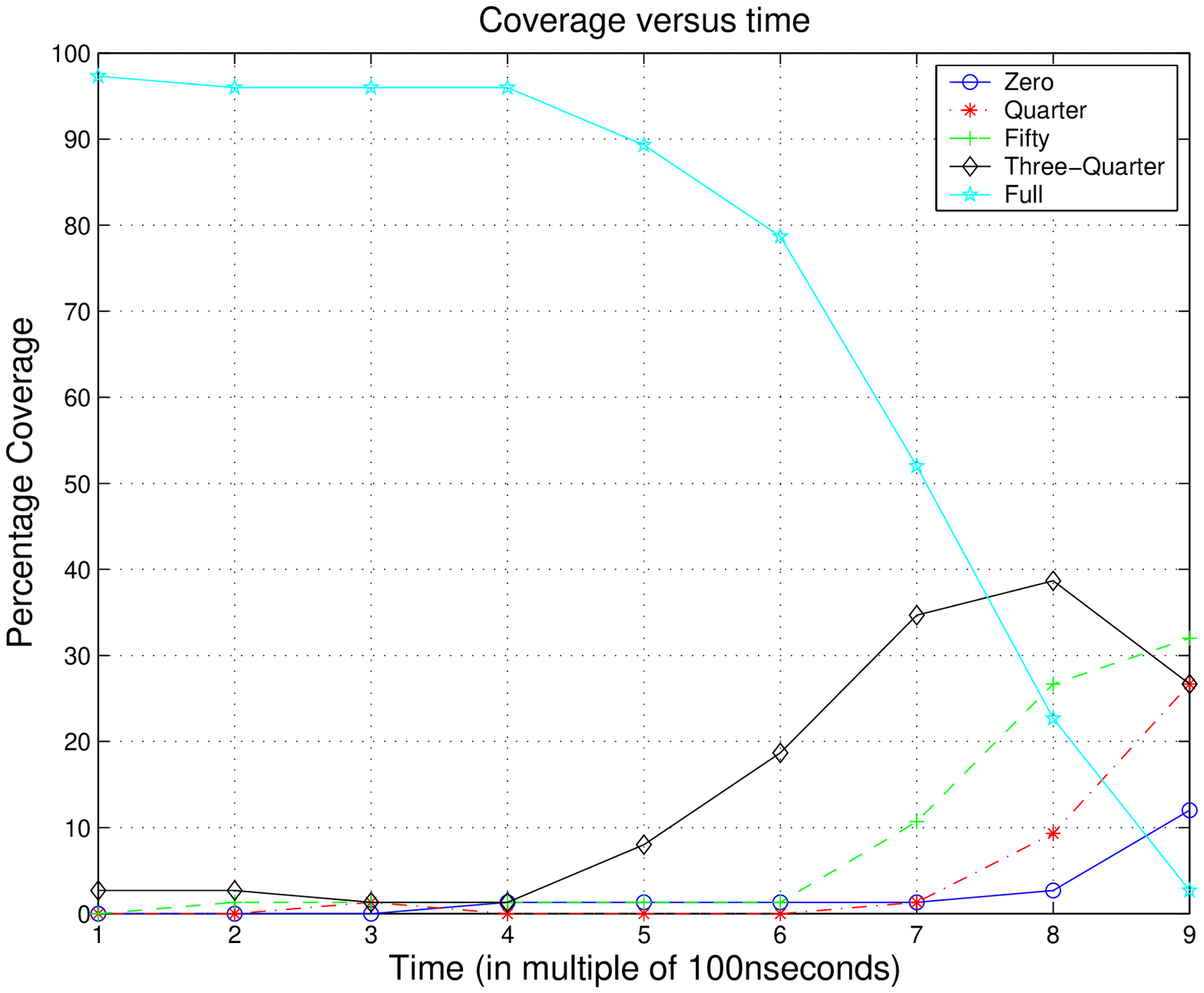,scale=0.4}}
} \mbox{ \subfigure[LS Manifold Coverage (Uneven Data
Generation)\label{localmaniuneven}]{\epsfig{file=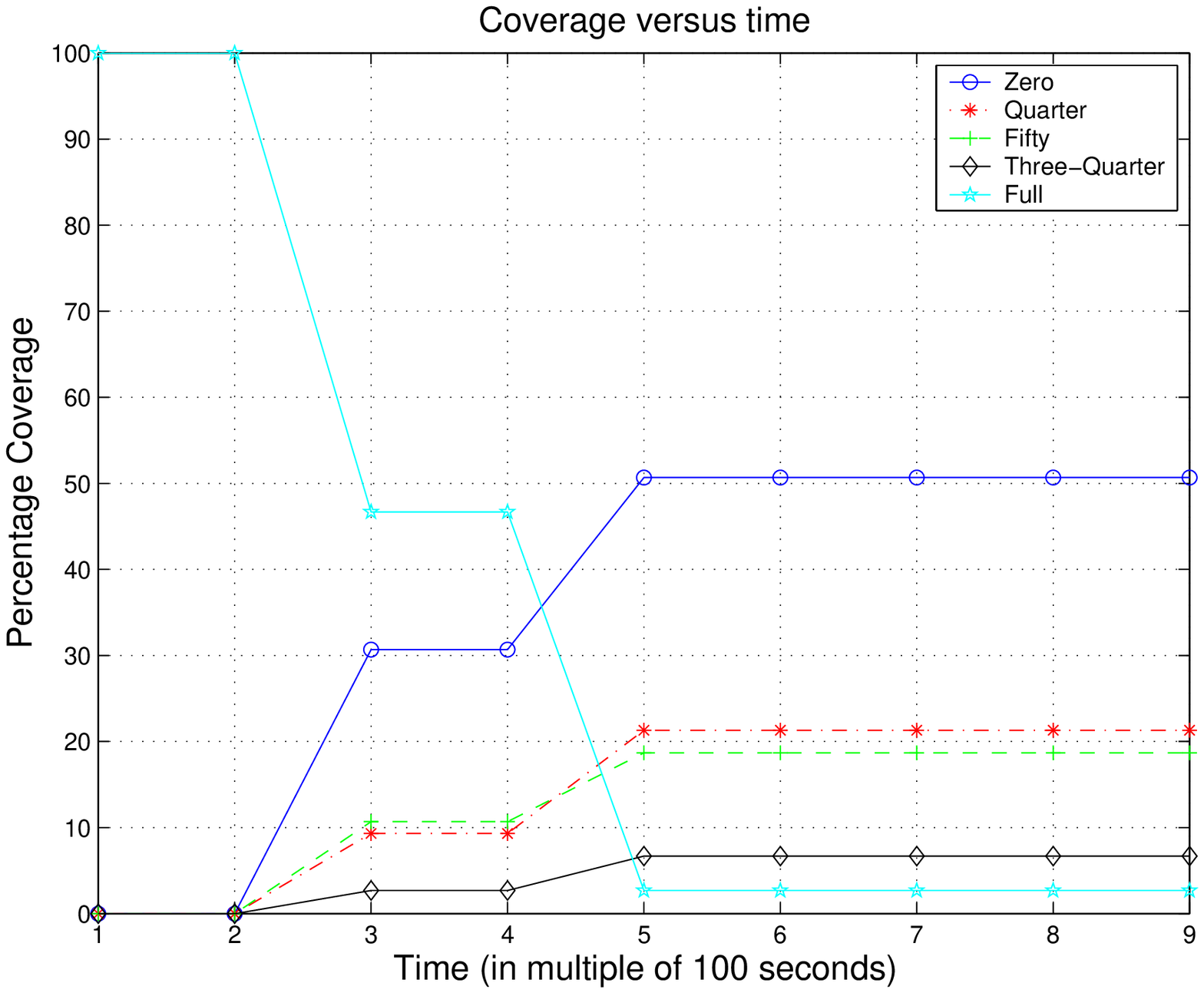,scale=0.4}}
\quad \subfigure[CBCS Manifold Coverage (Uneven Data
Generation)\label{clustmaniuneven}]{\epsfig{file=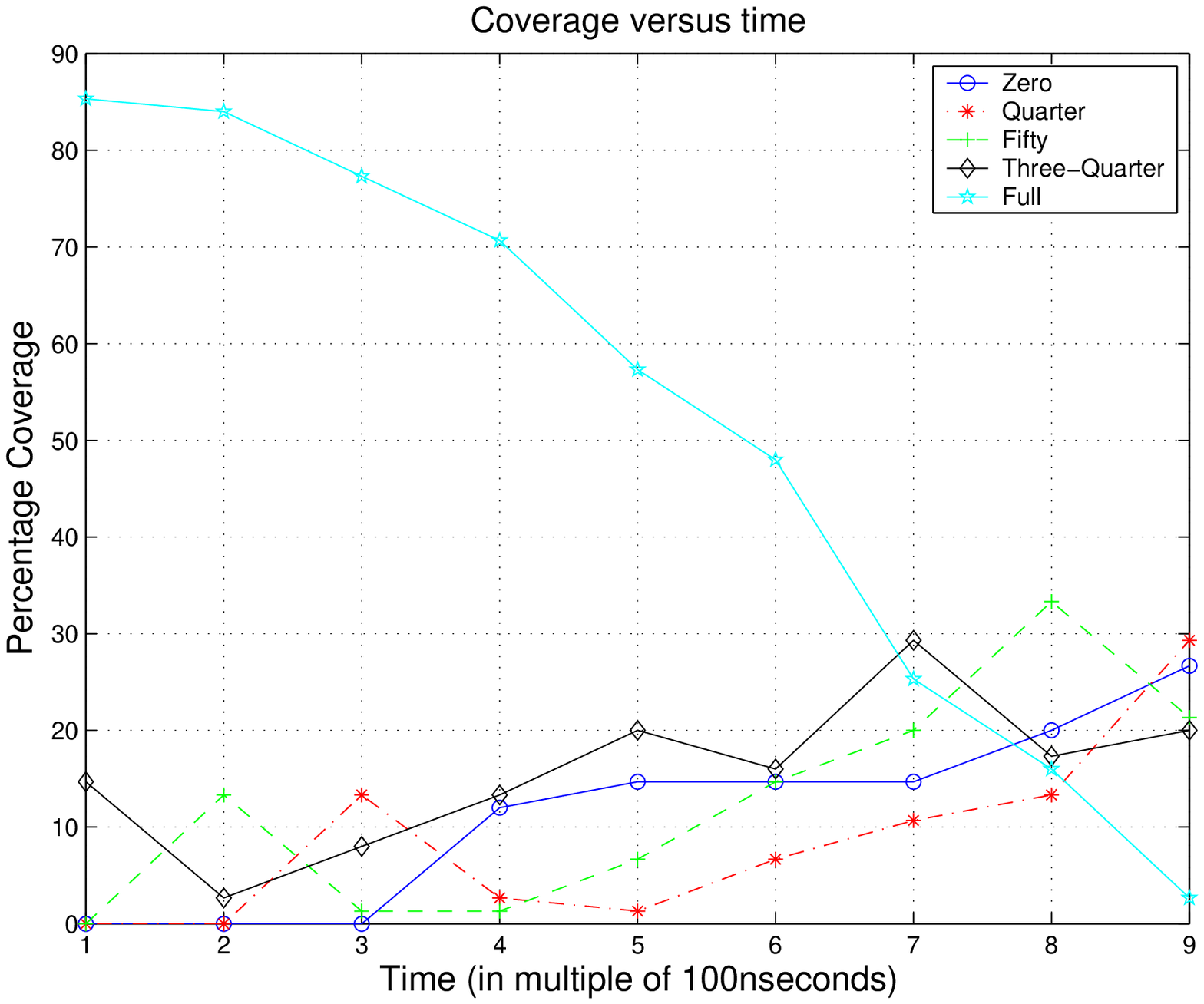,scale=0.4}}
}
\caption{Manifold Coverage}
\label{fig:mani2} 
\end{center}
\end{figure*}     

Similar trends are seen when considering Manifold Coverage
(Figure~\ref{fig:mani2}).  Each line represents the percentage of
zones with some specific coverage level: for example, the line
``quarter'' represents the percentage of zones where at least 25\%
of the sensors have storage space left.  One can clearly see that, in
the case of LS with even data generation (Figure~\ref{localmanieven})
the percentage of zones with full coverage is 100\% at 300 seconds,
whereas with uneven data generation it reduces to less than 50\%
within 300 seconds. In CBCS, at the same times, the coverage is around
96\% with the even data generation model and with the uneven data
model it is around 77\%. Note that, A CH stored more data than
individual sensor, therefore if the round time is very long, it might
happen that the given CH runs out of storage sooner than a sensor
storing its data locally. In LS, the percentage of dead zones (zones
with all sensors out of storage space) rises in two waves for the
uneven data model, reaching up to 30\% within 300 seconds and 50\% in
500 seconds
However, with CBCS, with the uneven data model, the
percentage of dead zones rises slowly and is below 30\% even at the
end of the simulation. 
In general, from these figures, one can see that the manifold coverage
changes are abrupt for local buffering. In contrast, collaborative
storage provide smooth degradation of coverage.  Moreover, the average
coverage is higher for collaborative storage due to the data
aggregation and load balancing ability.  
by transferring data from high activity zones to low
activity zones.

\subsection{Effect of the Aggregation Model}

One limitation of the aggregation model we have used so far is that
the required storage size under collaboration grows in direct
proportion to the number of sensors in the cluster; that is, the
storage consumed in a round is $\alpha N \cdot D$, where $\alpha$ is
the aggregation ratio, $N$ is the number of sensors and $D$ is the
data sample size.  Since the available storage ($N \cdot S$, where $S$
is the available storage per sensor) is also a function of the number
of sensors, storage is consumed at a rate ($frac{\alpha D}{S}$) which
is independent of the number of sensors present in the zone, assuming
perfect load balancing.  For most applications, this will not be the
case: the aggregated data necessary to describe the phenomenon in the
zone does not grow strictly proportionately to the number of sensors
and we expect storage lifetime to be longer in dense areas than in
sparse ones.  

To highlight the above effect, we consider the case of a
\textit{biased deployment} where sensors are deployed randomly but
with non-uniform density.
In addition to the aggregation model considered so far, we consider a
case where the CH upon receiving packets from its $N$ members, just
needs to store 1 packet. As an example if the aggregation function is
to store the average value of the $N$ samples (e.g. average
temperature reading). Clearly, in the second case, the size of the
aggregated data is independent of network density.  We now study how
these applications with different aggregation functions perform on top
of a biased deployment.  To model biased deployment, we consider 4
zones with 5,4,3,2 sensors respectively. In these simulations, the
round time was set to 10 seconds (CH selection happens every 10
seconds).

\begin{figure*}
\begin{center}
\mbox{
\subfigure[Aggregation ratio = 0.5: Coverage\label{biasedapt5}]{\epsfig{file=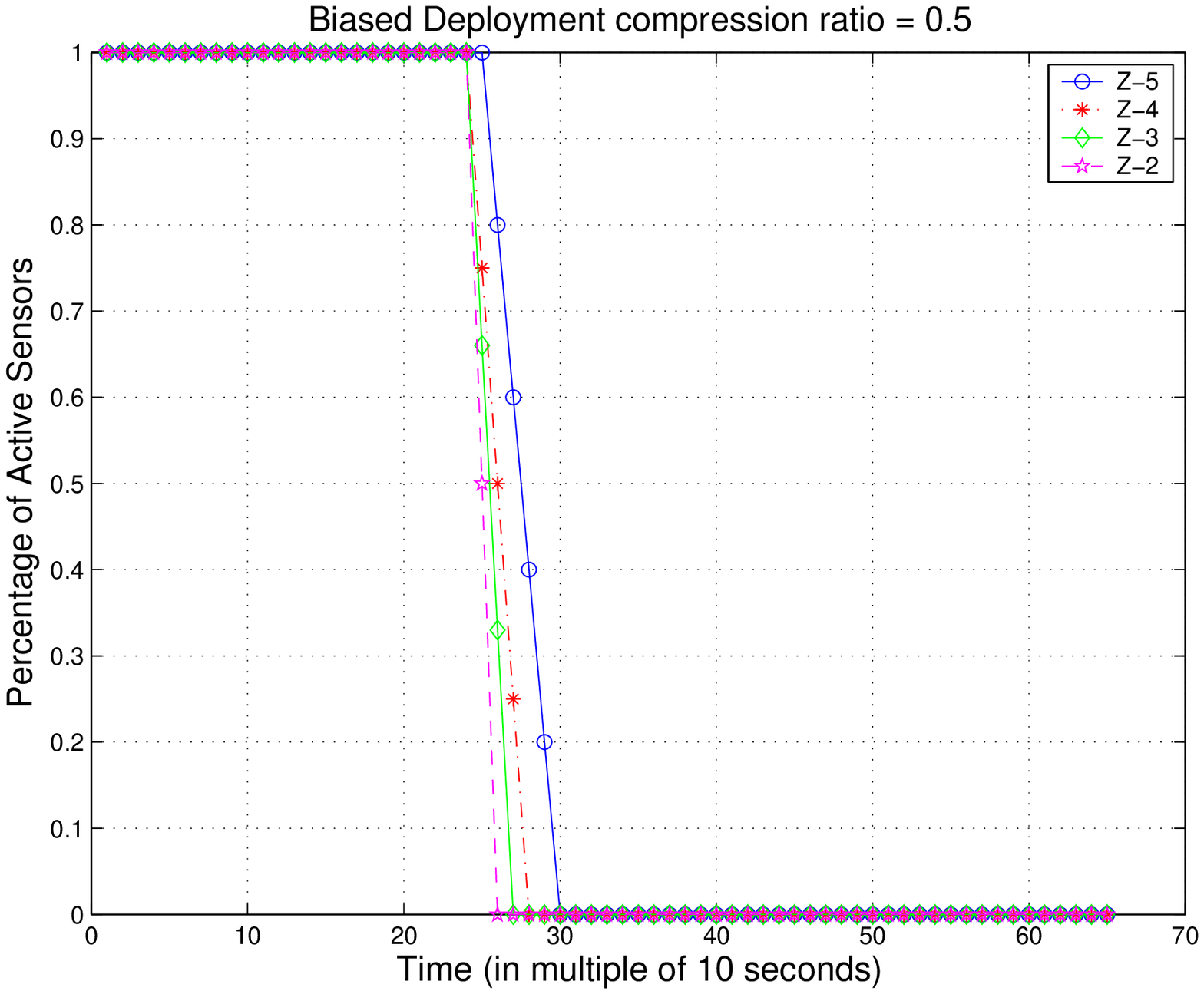,scale=0.4}} \quad
\subfigure[Aggregation Ratio = $1/N$: Coverage \label{biasedovern}]{\epsfig{file=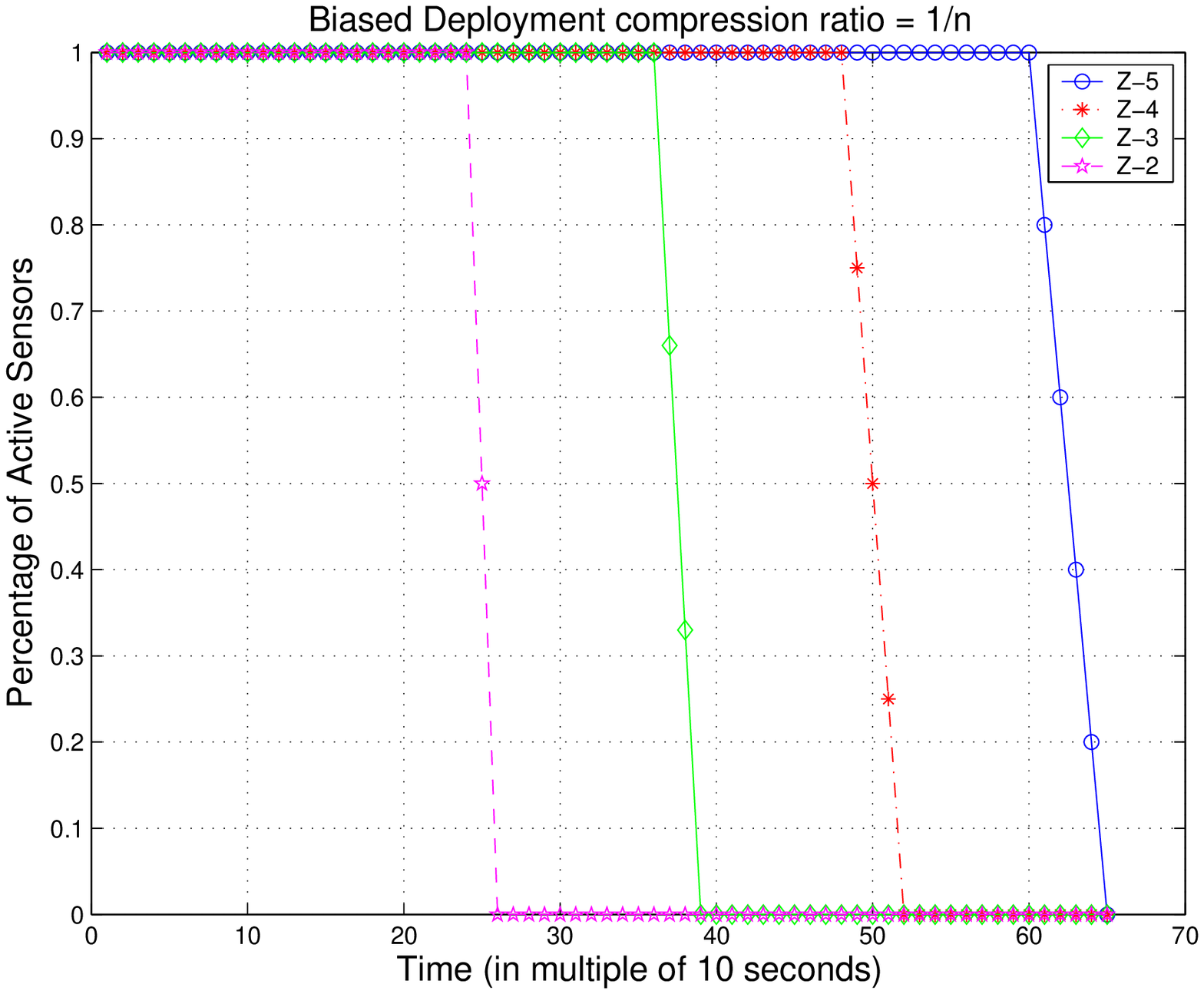,scale=0.4}}
}
\caption{Biased Deployment versus coverage study.}
\label{fig:biased} 
\end{center}
\end{figure*}

In Figure~\ref{fig:biased}, the X-axis shows time (in multiple of 10
seconds), whereas the Y-axis shows the percentage of coverage sensors
within a given zone. As described earlier we considered 4 zones for
this study and each line in the Figure~\ref{fig:biased} represents a
particular zone. For example line Z-5 stands for a zone with 5 sensors
in it and Z-2 denotes the zone with 2 sensors in it and so on.  As
shown in Figure~\ref{biasedapt5}, when the aggregation ratio is a
constant (0.5), all the zones provide coverage for almost same
duration. However, in the second case, as shown if
Figure~\ref{biasedovern}, coverage is directly proportional to the
network density, higher the density, longer the coverage.

The sensor network coverage from a storage management perspective
depends on the event generate rate, the aggregation properties as well
as the available storage.  If the aggregated data size is independent
of the number of sensors (or grows slowly with it), the density of the
zone correlates with the availability of storage resources.  Thus,
both the availability of storage resources as well as the consumption
of them may vary within a sensor network.  This argues for the need of
load-balancing across zones to provide long network lifetime and
effective coverage.  This is a topic of future research.

%% file: conclude.tex
\section{Conclusion and Future Work} \label{conclude}

In this paper, we considered the problem of storage management in
sensor networks where the data is not continuously reported in
real-time and must therefore be stored within the network.
Collaborative storage is a promising approach for storage management
because it enables the use of spatial data aggregation between
neighboring sensors to compress the stored data and optimize the
storage use.  Collaborative storage also allows load balancing of the
storage space to allow the network to maximize the time before data
loss due to insufficient memory.  Collaborative storage
results in lower time to transfer the data to the observer during the
reach-back stage and better binary and manifold coverage than a simple
local buffering approach.  Finally, we explored the use of
coordination to cut down on redundancy at the source sensors,
resulting in an improved version of both local storage and
collaborative storage.

  While collaborative storage reduces the energy required for storage,
it requires additional communication.  Using current technologies,
collaborative storage requires more energy than local buffering.
Network effectiveness is bound both by storage availability (to allow
continued storage of collected data) as well as energy.  Thus,
protocol designers must be careful to balance these constraints: if
the network is energy constrained, but has abundant storage, local
storage is most efficient from an energy perspective.  Alternatively,
if the network is storage constrained, collaborative storage is most
effective from a storage perspective.  When the network is constrained
by both, a combination of the two approaches would probably perform
best.

%% file: future-work.tex

As part of our future research, 
we would like to implement  these protocols on real sensor hardware
platforms such as the Berkeley motes.
Furthermore, in this study we consider all events to be of the
same importance, and thus they are
stored with the same compression ratio (resolution).  In our future
research, we will
explore the protocol space wherein different events are stored with different 
resolutions (important events are stored in detail whereas unimportant events
are stored with a coarser granularity).